\documentclass[letter]{aa}
\usepackage{graphicx}
\usepackage{txfonts}

\def\RpRs{\hbox{$R_{\rm p}/R_\star$}}
\def\Tmid{\hbox{$T_{\rm mid}$}}

\def\chisq{\mbox{$\chi^2$}}

\usepackage[colorlinks=true,citecolor=blue]{hyperref}
\usepackage{natbib}
\begin{document}

   \title{A feature-rich transmission spectrum for WASP-127b}
   \subtitle{Cloud-free skies for the puffiest known super-Neptune?}

   \author{E. Palle\inst{1,2}
          \and
          G. Chen\inst{1,2,3}
          \and
          J. Prieto-Arranz\inst{1,2}
          \and
          G. Nowak\inst{1,2}
          \and
          F. Murgas\inst{1,2}
          \and
          L. Nortmann\inst{1,2}
          \and
          D. Pollacco\inst{4}
          \and
          K. Lam\inst{4}
          \and
          P. Montanes-Rodriguez\inst{1,2}
          \and
          H. Parviainen\inst{1,2}
          \and
          N. Casasayas-Barris\inst{1,2}
          }

   \institute{Instituto de Astrof\'{i}sica de Canarias, V\'{i}a L\'{a}ctea s/n, E-38205 La Laguna, Tenerife, Spain\\
         \email{epalle@iac.es}
         \and
             Departamento de Astrof\'{i}sica, Universidad de La Laguna, Spain
         \and
             Key Laboratory of Planetary Sciences, Purple Mountain Observatory, Chinese Academy of Sciences, Nanjing 210008, China
         \and
             Department of Physics, University of Warwick, Coventry CV4 7AL, UK}

   \date{Received Month 00, 2017; accepted Month 00, 2017}

% \abstract{}{}{}{}{} 
% 5 {} token are mandatory
 
  \abstract
%  % context heading (optional)
%  % {} leave it empty if necessary  
   {WASP-127b is one of the lowest density planets discovered to date. With a sub-Saturn mass ($M_{\rm p}=0.18 \pm 0.02 M_J$) and super-Jupiter radius ($R_{\rm p}= 1.37 \pm 0.04 R_J$), it orbits a bright G5 star, which is about to leave the main-sequence.}
%  % aims heading (mandatory)
   {We aim to explore WASP-127b's atmosphere in order to retrieve its main atmospheric components, and to find hints for its intriguing inflation and evolutionary history.}
%  % methods heading (mandatory)
   {We used the ALFOSC spectrograph at the NOT telescope to observe a low resolution ($R\sim330$, seeing limited) long-slit spectroscopic time series during a planetary transit, and present here the first transmission spectrum for WASP-127b.}
%  % results heading (mandatory)
   {We find the presence of a strong Rayleigh slope at blue wavelengths and a hint of Na absorption, although the quality of the data does not allow us to claim a detection. At redder wavelengths the absorption features of TiO and VO are the best explanation to fit the data.}
%  % conclusions heading (optional), leave it empty if necessary 
   {Although higher signal-to-noise ratio observations are needed to conclusively confirm the absorption features, WASP-127b seems to posses a cloud-free atmosphere and is one of the best targets to perform further characterization studies in the near future.}

   \keywords{Planetary systems --
             Planets and satellites: individual: WASP-127b --
             Planets and satellites: atmospheres --
             Techniques: spectroscopic}

   \maketitle
%
%-------------------------------------------------------------------

%------------------------------------------------------------------------
%  1. INTRODUCTION
%------------------------------------------------------------------------
\section{Introduction}
\label{sec:intro}

The atmospheres of exoplanets are a unique window to investigate the planetary chemistry, which can help improve our understanding of planetary interior properties and provide links to planet formation and migration histories \citep[e.g.,][]{2005AREPS..33..493G,2007ApJ...659.1661F,2010SSRv..152..423F,2011ApJ...743L..16O,2012ApJ...751L...7M,2014ApJ...794L..12M,2016SSRv..205..285M}. Transmission spectroscopy retrieves the absorption and scattering signatures from the atmosphere at the planetary day-night terminator region. These signatures are only imprinted on the stellar light when it is transmitted through the planetary atmosphere during a transit, and they can be extracted through the differential method when compared to out-of-transit measurements. Such studies have been carried out by many ground-based large telescope and space telescope, in a wide range of spectral resolutions \citep[e.g.,][]{2002ApJ...568..377C,2010Natur.465.1049S,2010Natur.468..669B,2016Natur.529...59S}, resulting in robust detections of Na, K, H$_2$O, CO, and scattering hazes (see the inventory listed in \citealt{2014PASA...31...43B} and \citealt{2016Natur.529...59S}). A recent HST+Spitzer survey led by \citet{2016Natur.529...59S} performed a comparative study on ten hot Jupiters covering $0.3-5 \mu m$. This diverse hot Jupiter sample reveals a continuum from clear to cloudy atmospheres, and suggests clouds/hazes as the cause of weakened spectral features.

As the investigated sample increases, it is fundamental to construct a spectral sequence for exoplanets, for a global picture of population characteristics and formation/evolution scenarios, as we have achieved for stars and brown dwarfs.  In the near-future the JWST will provide spectral resolutions at high SNR with a large wavelength coverage $0.6-28 \mu m$ that can distinguish among different atmospheric compositions. However, ground-based observations can also complement JWST by extending the wavelength range to $\lambda < 600 nm$, which is critical to examine spectral signatures arising from Rayleigh scattering, Na, or TiO/VO \citep{2014A&A...563A..41M,2016A&A...594A..65N,2017A&A...600A.138C,2017A&A...600L..11C}. The ideal starting point are low density planets, which are more likely to host extended atmospheric envelopes that can produce stronger transmission signals if cloud-free.

WASP-127b \citep{2017A&A...599A...3L}, with a mass of $0.18 \pm 0.02 M_J$ and a radius of $1.37 \pm 0.04 R_J$, is the puffiest, lowest density, planet discovered to date. It has an orbital period of 4.18 days, and orbits a bright parent star (V = 10.2), which makes it a very interesting object for atmospheric follow-up studies.

WASP-127b's host star is a G5 star which is at the end of the main-sequence phase and moving to the sub-giant branch \citep{2017A&A...599A...3L}. Moreover, the unusually large radius (compared to its sub-Saturn mass) cannot be explained by the standard coreless model \citep[e.g.,][]{2007ApJ...659.1661F}, and places it into the short-period Neptune desert, a region between Jovian and super-Earth planets with a lack of detected planets \citep{2012ApJS..201...15H,2016A&A...589A..75M}. Several inflation mechanisms have been proposed to explain this inflation, including tidal heating, enhanced atmospheric opacity, Ohmic heating, and/or re-inflation by host star when moving towards the RGB phase \citep{2010A&A...516A..64L,2010ApJ...714L.238B,2011ApJ...738....1B,2013MNRAS.430.1247L,2013ApJ...764..103R,2013ApJ...772...76S,2013ApJ...763...13W,2016ApJ...818....4L}, although no concluding observations have yet been established to favour one or the other. Therefore, the formation and evolution mechanisms of WASP-127b are very intriguing, given its transition size between these two classes of planets.

%++++++++++++++++++++++++++++++++++
%   Figure
%++++++++++++++++++++++++++++++++++
\begin{figure}[h!]
\centering
\includegraphics[width=\linewidth]{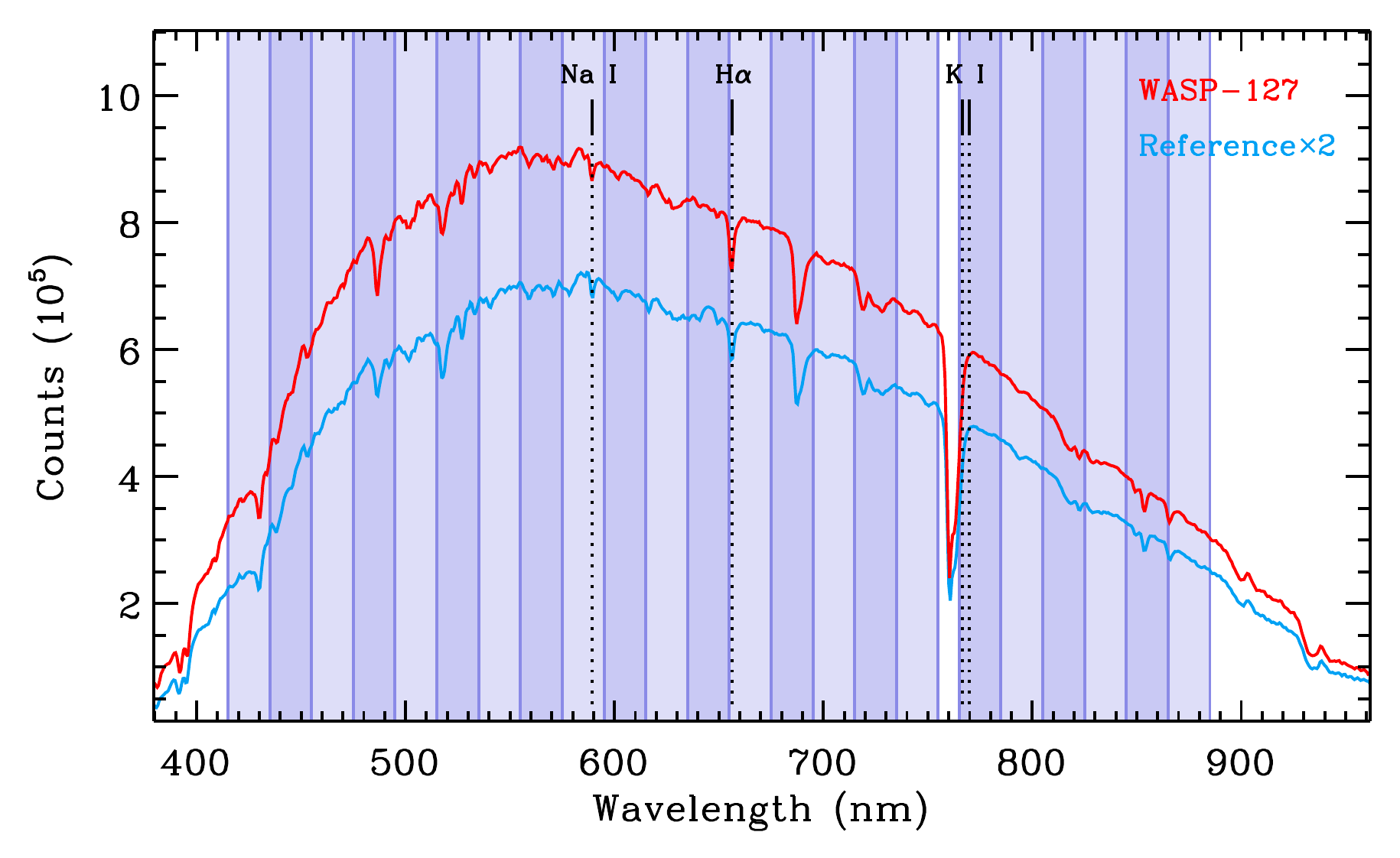}
\caption{Example stellar spectra of \object{WASP-127} (red) and the reference star (blue) obtained with the Grism \#4 of NOT/ALFOSC on the night of February 23, 2017. The color-shaded areas indicate the divided passbands that are used to create the spectroscopic light curves. Note how the  oxygen-A band region is excluded.\label{fig:Spectra}}
\end{figure}

%------------------------------------------------------------------------
%  2. OBSERVATIONS AND DATA REDUCTION
%------------------------------------------------------------------------
\section{Observations and data reduction}
\label{sec:data}

We observed one transit of \object{WASP-127b} on the night of February 23rd, 2017, using the Andalucia Faint Object Spectrograph and Camera (ALFOSC) mounted at the 2.5~m Nordic Optical Telescope (NOT) at ORM observatory. ALFOSC has a field of view of 6'.4x6'.4 and a 2048x2048 E2V detector with a pixel size of 0''.2.
The observation was carried out in the long-slit mode using a 40" wide slit to avoid flux losses, and placing both WASP-127 and a reference star simultaneously aligned into the slit. The reference star TYC 4916-897-1 is located 40''.5 away from WASP-127 and it is about one magnitude fainter (V = 11.2) over the observed spectral range. Grism \#4 was used covering simultaneously the spectral range from 320-960 $nm$. Observations started at 23:45 UT and ended at 05:36 UT, resulting in a time series of 746 spectra. Exposure times were set to 20s. The transit of WASP-127b ($T_{14}$) started at 00:19 UT and ended at 04:38 UT, resulting in 554 spectra taken within transit. The night was clear, with a relatively stable seeing of around 0".5 during the full observation. The airmass changed from 1.35 to 1.19, then to 2.45.

Data reduction was carried out using the approach outlined in \citet{chen2016a,2017A&A...600A.138C} for similar OSIRIS long-slit data taken with the GTC. The one-dimensional spectra (see Figure~\ref{fig:Spectra}) were extracted using the optimal extraction algorithm
\citep{1986PASP...98..609H} with an aperture diameter of 13 pixels, which minimized
the scatter for the white-color light curves created from
various trial aperture sizes. The time stamp was centered on mid-exposure
and converted into the Barycentric Dynamical Time
standard \citep[BJD$_\mathrm{TDB}$;][]{2010PASP..122..935E}. Misalignment between the target and reference stars in the wavelength solutions
and any spectral drifts were corrected. Then the requested wavelength
range of a given pass-band was converted to a pixel range,
and the flux was summed to generate the time series.

\begin{figure}
\centering
\includegraphics[width=0.8\linewidth]{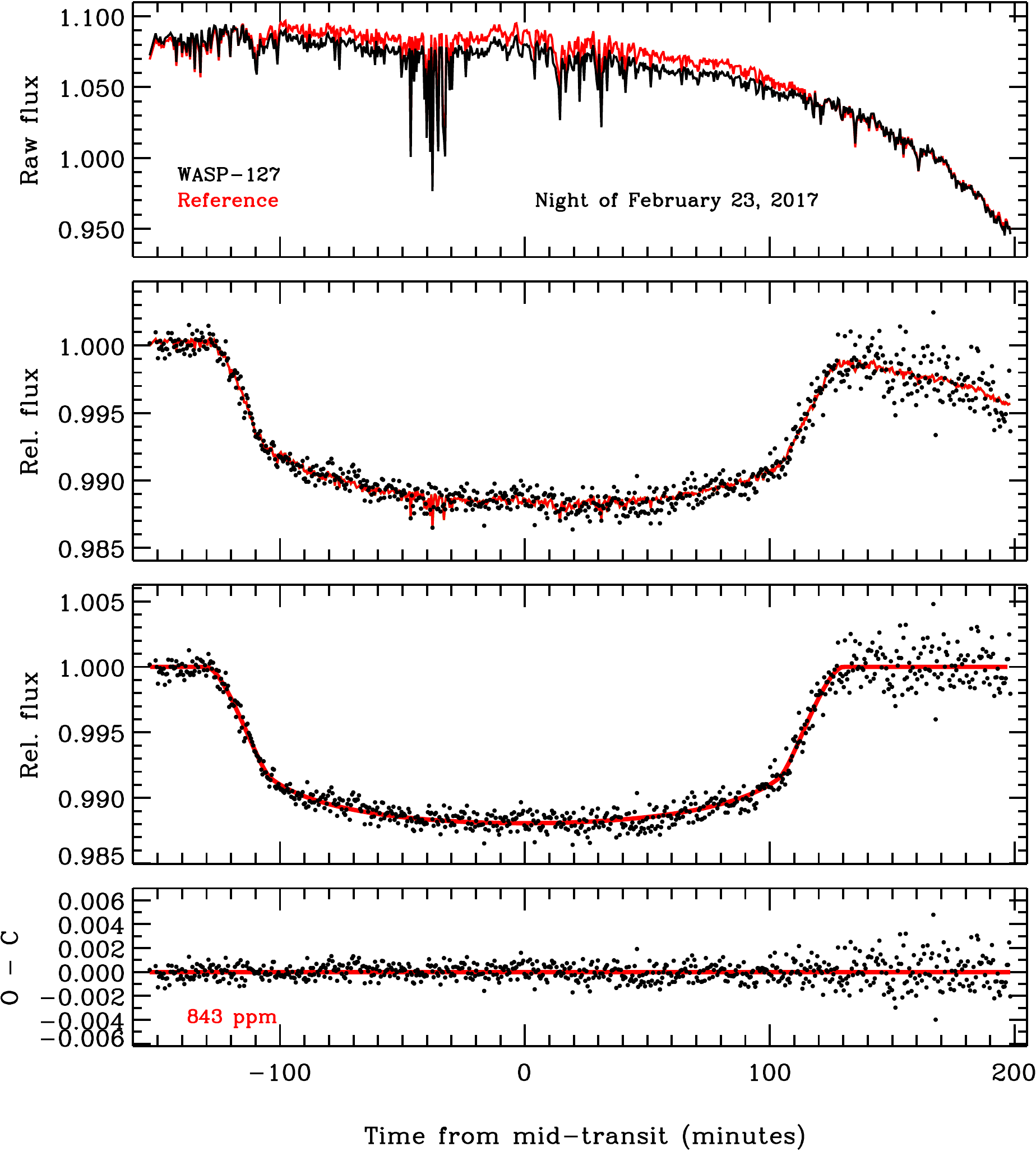}
\caption{Panels from top to bottom: (1) raw flux of WASP-127 (black line) and the reference star (red line) obtained with ALFOSC at NOT; (2) relative flux between the WASP-127 and the reference star (black dots) and the best-fitting combined model (red line), (3) same as in panel 2, but detrended, (4) best-fitting light-curve residuals. \label{fig:WhiteLC}}
\end{figure}

A broad-band (white-color) light curve was integrated from 395 nm to 945 nm,
excluding the range of 755–765 nm to eliminate the noise introduced
by the oxygen-A band \citep{parvi}, and used to derive the transit parameters in Figure~\ref{fig:WhiteLC}. Moreover several narrow band light curves were constructed to study the wavelength-dependence of the transit depth and derive the transmission spectrum (see Figure~\ref{fig:Spectra} for the band ranges).

%++++++++++++++++++++++++++++++++++
%   Figure
%++++++++++++++++++++++++++++++++++

%------------------------------------------------------------------------
%  3. ANALYSIS
%------------------------------------------------------------------------
\section{Light-curve analysis}
\label{sec:analysis}

The light-curve data were modeled in the approach detailed in \citet{chen2016a,2017A&A...600A.138C}. In brief, the light-curve model contains two multiplicative components. One component describes the astrophysical signal, which adopts the analytic transit model $\mathcal{T}(p)$ proposed by \citet{2002ApJ...580L.171M}. The other component describes the systematics of telluric or instrumental origins in a fully parametric form or in a semi-empirical form, which is designated as the baseline model $\mathcal{B}(c_i)$. 

The transit model $\mathcal{T}(p)$ was parameterized as orbital period $P$, inclination $i$, scaled semi-major axis $a/R_\star$, planet-to-star radius ratio $R_{\rm p}/R_\star$, mid-transit time $T_{\rm mid}$, and limb-darkening coefficients $u_i$, where a circular orbit was assumed. The orbital period $P$ was fixed to 4.178062~days as reported by \citet{2017A&A...599A...3L}. A quadratic limb-darkening law was adopted and conservatively constrained by Gaussian priors of width $\sigma=0.1$, whose central values were calculated from the ATLAS atmosphere models following \citet{2015MNRAS.450.1879E} with stellar parameters $T_{\rm eff}=5750$~K, $\log g=3.9$, and $[\rm{Fe/H}]=-0.18$.

The baseline model $\mathcal{B}(c_i)$ consisted of a selected combination of auxiliary state vectors, including spectral and spatial position drifts ($x$, $y$), spectra's full width at half maximum (FWHM) in the spatial direction ($s_y$), airmass ($z$), and time sequence ($t$). The Bayesian information criterion \citep[BIC;][]{Schwarz1978} was used to find the baseline model that can best remove the systematics. For the white-color light-curve, the model
\begin{align}
\mathcal{B_\mathrm{w}} = & ~c_0+c_1s_y+c_2z
\end{align}
gave the lowest BIC value. The second best model yields a value of $\Delta\mathrm{BIC}=3.3$ higher. For the spectroscopic light-curves, the model was chosen in a semi-empirical form:
\begin{align}
\mathcal{B_\mathrm{spec}(\lambda)} = & ~\mathcal{S_\mathrm{w}}\times\big(c_0+c_1s_y(\lambda)+c_2t+c_3t^2\big),
\end{align}
which inherited a common-mode component $\mathcal{S_\mathrm{w}}$ determined from the white-color light-curve. The common-mode systematics $\mathcal{S_\mathrm{w}}$ were derived after dividing the white-color light-curve by the best-fitting transit model $\mathcal{T}(p)$.

The Transit Analysis Package \citep[TAP;][]{2012AdAst2012E..30G}, customized for our purposes, was employed to perform the Markov chain Monte Carlo analysis. The correlated noise was taken into account by the wavelet-based likelihood function proposed by \citet{2009ApJ...704...51C}. The overall transit parameters were determined from the white-color light-curve, whose best-fitting values and associated uncertainties were calculated as the median and 1$\sigma$ percentiles of the posterior probability distributions and listed in Table~\ref{tab:tran_param}. For the spectroscopic light-curves, only the planet-to-star radius ratio $R_{\rm p}/R_\star$, the limb-darkening coefficients $u_i$, and the baseline coefficients $c_i$ were fit, while the other transit parameters were fixed to the ones determined from the white-color light-curve. The wavelength-dependent radius ratios are presented in Table~\ref{tab:transpec}. The white-color and spectroscopic light-curves are shown in Fig.~\ref{fig:WhiteLC} and \ref{fig:SpecLC}, respectively.

%++++++++++++++++++++++++++++++++++
%   Figure
%++++++++++++++++++++++++++++++++++
\begin{figure*}
\centering
\includegraphics[width=\linewidth]{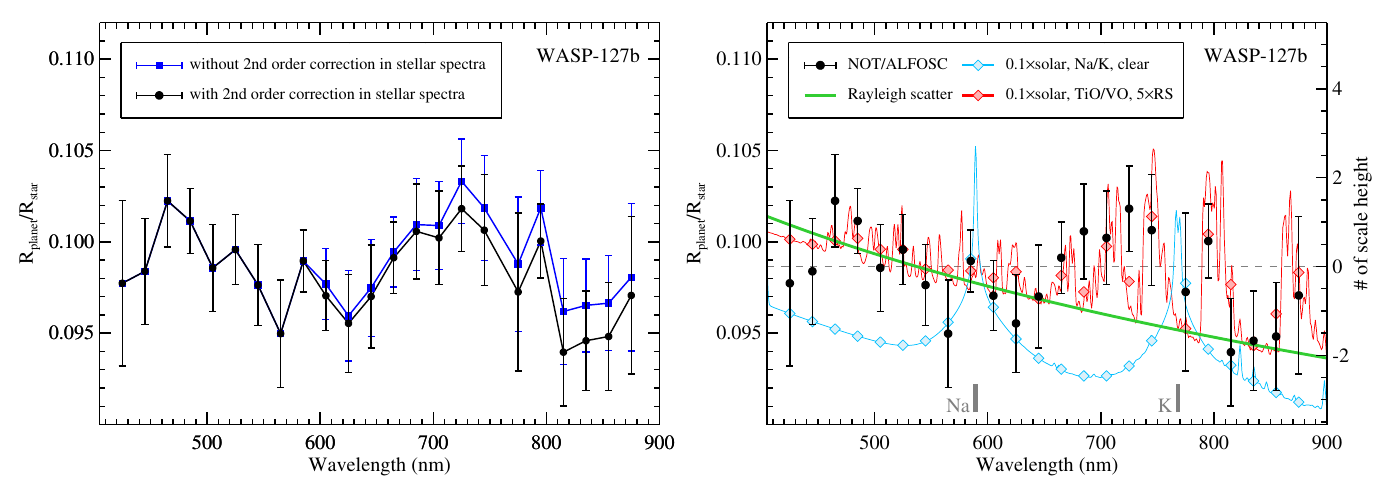}
\caption{The NOT/ALFOSC transmission spectrum of \object{WASP-127b}, using bins 20 $nm$ wide in wavelength. Error bars are $\pm 1\sigma$ errors. The left panel shows the transmission spectra with (black) or without (blue) the second order correction in the stellar spectra. The right panel shows two 0.1$\times$solar atmospheric models computed using \texttt{Exo-Transmit} \citep[][red model: with TiO/VO but without Na/K, 5$\times$ Rayleigh scattering, sky-blue model: with Na/K but without TiO/VO, clear]{2016arXiv161103871K}, and a pure Rayleigh slope, together with the corrected transmission spectrum.
\label{fig:Transpec}}
\end{figure*}

%++++++++++++++++++++++++++++++++++
%   Table
%++++++++++++++++++++++++++++++++++
\begin{table}
     \small
     \centering
     \caption{Derived system parameters for white light curve analysis}
     \label{tab:tran_param}
     \begin{tabular}{lr}
     \hline\hline\noalign{\smallskip}
     Parameter & Value\\\noalign{\smallskip}
     \hline\noalign{\smallskip}
     $P$ [days]   & 4.178062 (fixed) \\\noalign{\smallskip}
     $e$   & 0 (fixed) \\\noalign{\smallskip}
     \Tmid\ [$\mathrm{BJD}_\mathrm{TDB}$] & 2457808.60283 $\pm$ 0.00031\\\noalign{\smallskip}
     $i$ [$^\circ$] & 88.2 $^{+1.1}_{-0.9}$ \\\noalign{\smallskip}
     $a/R_\star$ &  7.95 $^{+0.19}_{-0.27}$ \\\noalign{\smallskip}
     \RpRs & 0.1004 $\pm$ 0.0014 \\\noalign{\smallskip}
     $u_1$ & 0.365 $\pm$ 0.057 \\\noalign{\smallskip}
     $u_2$ & 0.258 $^{+0.88}_{-0.85}$ \\\noalign{\smallskip}
    \hline\noalign{\smallskip}
    \end{tabular}
\end{table}

%------------------------------------------------------------------------
%  4. DISCUSSION
%------------------------------------------------------------------------
\section{Results and Discussion}
\label{sec:discuss}

%------------------------------------------------------------------------
\subsection{Second order contamination}

When using the grism \#4 with ALFOSC, second order contamination can be present due to the overlap in the detector of different diffraction orders 
\citep{2007AN....328..948S}. To check this issue, on March 21st 2017, we performed consecutive observations of WASP-127 with the grism \#4, with and without the second order blocking filters \#101 (GG475) and \#102 (OG515). We find that for WASP-127, second order contamination of the stellar flux appears at 1\% level at 655 $nm$ and rises nearly monotonically reaching 10\% at 900 $nm$. Following the approach of \citet{2007AN....328..948S}, we were able to directly remove the second order component of the blue light from the first order stellar spectra, and then to derive a new transmission spectrum. As shown in Fig.~\ref{fig:Transpec}, this correction makes the transit depths slightly smaller at red wavelengths ($\lambda\gtrsim 600$~nm), which agree with the original ones well within the error bars and still show the same relative spectral shape.

%++++++++++++++++++++++++++++++++++
%   Table
%++++++++++++++++++++++++++++++++++
\begin{table}[h!]
     \small
     \centering
     \caption{Goodness of fit for different atmospheric models}
     \label{tab:vsmodel}
     \begin{tabular}{lcccc}
     \hline\hline\noalign{\smallskip}
     Model & \multicolumn{2}{c}{415--885~nm} & \multicolumn{2}{c}{415--655~nm} \\\noalign{\smallskip}
        & $\chi^2$ & $P(\chi^2)$ & $\chi^2$ & $P(\chi^2)$ \\\noalign{\smallskip}
     \hline\noalign{\smallskip}
       Pure Rayleigh Scattering (RS)                & 22.11 & 0.453  &  4.89  & 0.936 \\\noalign{\smallskip}
       Flat                                         & 18.83 & 0.656  &  8.43  & 0.645 \\\noalign{\smallskip}
     \hline\noalign{\smallskip}                                    
       1$\times$solar, Na/K, clear                  & 40.41 & 0.010  & 26.59  & 0.005 \\\noalign{\smallskip}
       0.1$\times$solar, Na/K, clear                & 30.64 & 0.104  & 14.71  & 0.196 \\\noalign{\smallskip}
       1$\times$solar, Na/K/TiO/VO, 5$\times$RS     & 24.97 & 0.299  & 10.61  & 0.477 \\\noalign{\smallskip}
       0.1$\times$solar, Na/K/TiO/VO, clear         & 23.95 & 0.350  & 13.40  & 0.268 \\\noalign{\smallskip}
       0.1$\times$solar, Na/K, 5$\times$RS          & 23.29 & 0.386  &  5.79  & 0.887 \\\noalign{\smallskip}
       0.1$\times$solar, Na/K/TiO/VO, 5$\times$RS   & 13.60 & 0.915  &  5.51  & 0.904 \\\noalign{\smallskip}
       0.1$\times$solar, TiO/VO, 5$\times$RS        & {\bf 11.73} & {\bf 0.963} & {\bf 4.45} & {\bf 0.955} \\\noalign{\smallskip}
    \hline\noalign{\smallskip}
    \end{tabular}
\end{table}

\subsection{Transmission spectrum}
\label{sec:transpec}

To interpret the transmission spectrum of WASP-127b, a series of atmospheric models with an isothermal temperature structure were generated using the \texttt{Exo-Transmit} code \citep{2016arXiv161103871K}. Various metallicities, chemical compositions (with or without the presence of Na, K, TiO, VO), and weather conditions (clear,  hazy, or cloudy) were considered. We also analytically calculated a pure Rayleigh scattering model following the approach of \citet{2008A&A...485..865L}, and used a simple flat straight line to represent the gray absorbing clouds.

It is clear from Fig.~\ref{fig:Transpec} that the transmission spectrum is not flat, but on the contrary it has strong spectral features. 
It is not surprising that we can detect spectral features even using a relatively small aperture telescope, given that one atmospheric scale height, $H$, of WASP-127b corresponds to approximately 2500 $km$ (equivalent to a signal of $510~ppm$) assuming an H-He atmosphere, and that the amplitude of a given spectral signature can typically achieve about $~5H$ \citep[e.g.,][]{2010eapp.book.....S}.

At the bluer wavelengths, the spectrum shows a decreasing slope with $\lambda$, which seems to indicate the presence of Rayleigh scattering. A hint of Na absorption is seen (although statistically insignificant), with the band centered on the Na doublet presenting a larger $R_{\rm p}/R_\star$ value than the surrounding bands. Unfortunately, the analysis of narrower pass bands around Na did not provide more information but increasing noise (not shown). No K absorption is seen. Toward the red, strong absorptions from TiO and VO molecules seem to dominate the spectral shape. 

Fitting the different models to the whole spectral range (415-885 $nm$) or the blue spectral range free of second order (415--655 $nm$), the one with the minimum $\chisq$ is always precisely the model including only TiO/VO, and with an enhanced Rayleigh slope indicative of some haze in the atmosphere (see Figure~\ref{fig:Transpec}, and Table~\ref{tab:vsmodel} for $\chisq$ fitting results). We find that the best fitting models are metal poor, which is interesting because the host star is also metal poor ([Fe/H]= $-0.18 \pm 0.06$).

Given the relatively cool equilibrium temperature of WASP-127b \citep[$T_{\rm eq}=1400$~K;][]{2017A&A...599A...3L}, the tentative inference of the TiO/VO molecules is somewhat unexpected and intriguing. For planets with equilibrium temperatures lower than $\sim$1900~K, TiO could be cold trapped in the deep atmospheric layers when the temperature-pressure profile crosses the condensation curve \citep{2009ApJ...699..564S,2013A&A...558A..91P,2016ApJ...828...22P}. Several other possibilities could also account for TiO/VO's absence in the upper atmosphere \citep{2009ApJ...699.1487S}. Until now only two very hot Jupiters, that is WASP-121b \citep[$T_{\rm eq}=2400$~K;][]{2016ApJ...822L...4E} and WASP-48b \citep[$T_{\rm eq}=1956$~K;][]{murgas2017}, have shown evidence of TiO/VO in the transmission spectrum. If the presence of TiO/VO were true in WASP-127b's relatively ``cool'' atmosphere, one possible scenario to avoid the cold trap could be that the stellar irradiation is directly deposited into WASP-127b's deep interior which thereby changes the deep temperature profile \citep[e.g.,][]{2010ApJ...719.1421P,2010ApJ...714L.238B,2011ApJ...738....1B,2012ApJ...757...47H,2013ApJ...764..103R,2013ApJ...772...76S,2016ApJ...818....4L}.

%------------------------------------------------------------------------
%  5. CONCLUSIONS
%------------------------------------------------------------------------
\section{Conclusions}
\label{sec:conclusions}

We have observed one transit of \object{WASP-127b}, an inflated, sub-Neptune mass planet. Because of its low density, the observed atmospheric scale height signals are large, and even with the NOT telescope we could retrieve its transmission spectrum. After considering the possible effects of second order contamination in the spectra, the spectrum shows the presence of a strong Rayleigh-like slope at blue wavelengths and a hint of Na absorption, although the quality of the data does not allow us to claim a detection. At redder wavelengths the absorption features of TiO and VO are the best explanation to fit the observed data. While the SNR is small, these findings are enough to conclude that the atmosphere of WASP-127b is either completely or partially cloud-free.

The brightness of its host star, a close-by comparison star, its extraordinary inflation, and its short period, all contribute to make WASP-127b a prime target for further followup with ground- and space-based facilities, including the JWST, which will be able to confirm our findings and extend them into the infrared regime. Finding the physical mechanism(s) responsible for this inflation will help us understand how this type of planets evolve and how their fate is tied to that of their host star.

\begin{acknowledgements}
    
    This article is based on observations made in the Observatorios de Canarias del IAC with the NOT telescope operated on the island of La Palma by the NOTSA in the Observatorio del Roque de los Muchachos (ORM).

    This work is partly financed by the Spanish MINECO through grants ESP2013-48391-C4-2-R, and ESP2014-57495-C2-1-R. 
    G.C. acknowledges the support by the National NSF of China (Grant No. 11503088) and the Nat. Sci. Found. of Jiangsu Province (Grant No. BK20151051). DLP is supported by the UK's STFC and a Royal Society Wolfson Merit award.
\end{acknowledgements}

\bibliographystyle{aa} % style aa.bst
\bibliography{ref_db} % your references Yourfile.bib

%------------------------------------------------------------------------
%  Appendix
%------------------------------------------------------------------------

%\Online
%\onecolumn

\begin{appendix}

\section{Appendix A: Spectro-photometric data}

\begin{figure}[h!]
\centering
\includegraphics[width=0.49\textwidth]{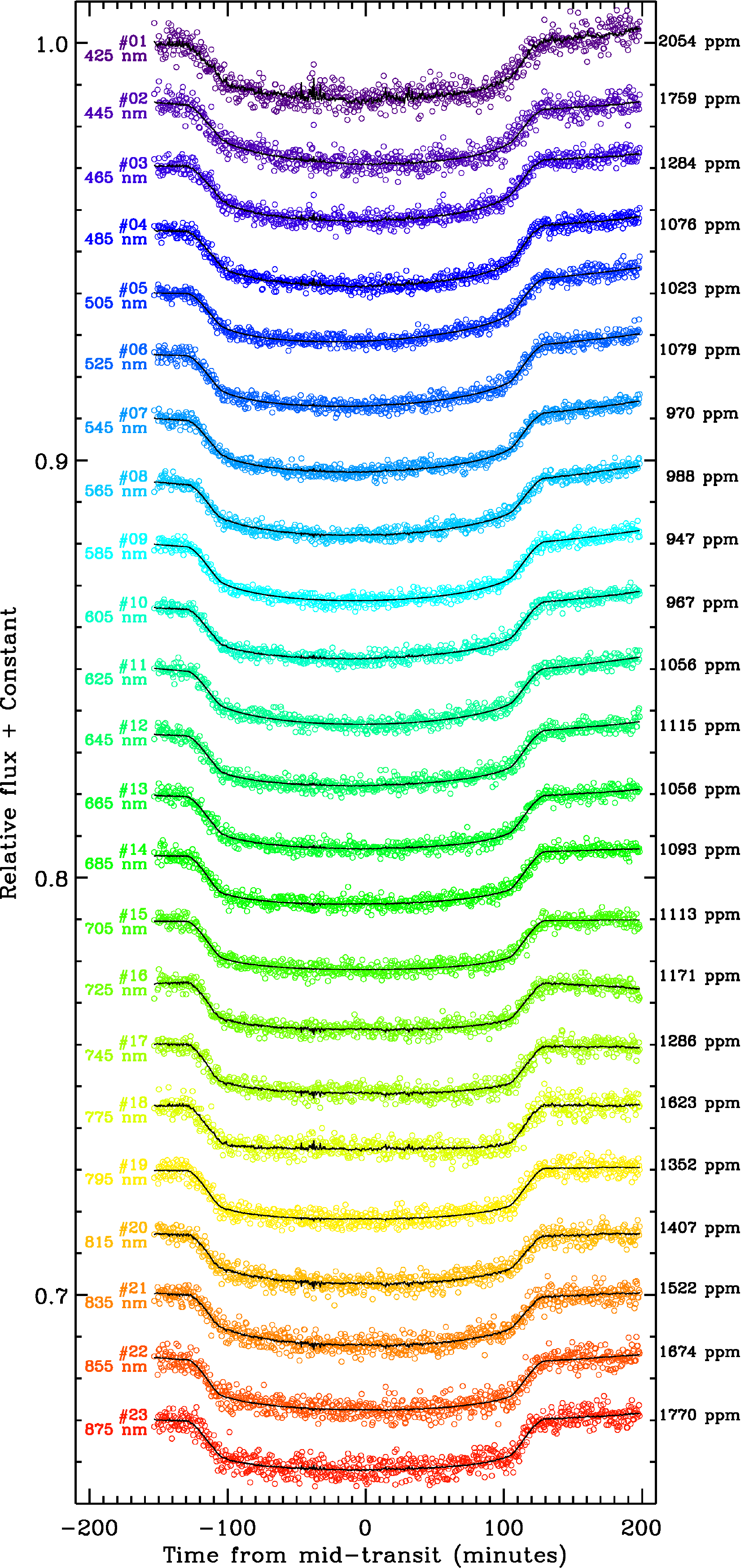}
\caption{Spectroscopic light-curves of WASP-127 and the best-fitting models after the common-mode systematics have been removed. \label{fig:SpecLC}}
\end{figure}

Observed color light curves are shown in Figure~\ref{fig:SpecLC} and the derived transit depths at each spectral pass band are given here in Table~\ref{tab:transpec}.

%++++++++++++++++++++++++++++++++++
%   Figure
%++++++++++++++++++++++++++++++++++

%++++++++++++++++++++++++++++++++++
%   Table
%++++++++++++++++++++++++++++++++++
\begin{table}[h!]
     \small
     \centering
     \caption{Transmission spectrum values obtained with NOT/ALFOSC}
     \label{tab:transpec}
     \begin{tabular}{ccccc}
     \hline\hline\noalign{\smallskip}
     \# & \multicolumn{2}{c}{Wavelength~(nm)} & \RpRs & \RpRs \\\noalign{\smallskip}
        & Center & Width & (With 2nd order) & (Corrected) \\\noalign{\smallskip}
     \hline\noalign{\smallskip}
       1 &          425 &  20  & 0.0977 $\pm$ 0.0045 & 0.0977 $\pm$ 0.0045 \\\noalign{\smallskip}
       2 &          445 &  20  & 0.0984 $\pm$ 0.0029 & 0.0984 $\pm$ 0.0029 \\\noalign{\smallskip}
       3 &          465 &  20  & 0.1022 $\pm$ 0.0025 & 0.1022 $\pm$ 0.0025 \\\noalign{\smallskip}
       4 &          485 &  20  & 0.1011 $\pm$ 0.0018 & 0.1011 $\pm$ 0.0018 \\\noalign{\smallskip}
       5 &          505 &  20  & 0.0986 $\pm$ 0.0024 & 0.0986 $\pm$ 0.0024 \\\noalign{\smallskip}
       6 &          525 &  20  & 0.0996 $\pm$ 0.0019 & 0.0996 $\pm$ 0.0019 \\\noalign{\smallskip}
       7 &          545 &  20  & 0.0976 $\pm$ 0.0022 & 0.0976 $\pm$ 0.0022 \\\noalign{\smallskip}
       8 &          565 &  20  & 0.0950 $\pm$ 0.0029 & 0.0950 $\pm$ 0.0029 \\\noalign{\smallskip}
       9 &          585 &  20  & 0.0990 $\pm$ 0.0017 & 0.0990 $\pm$ 0.0017 \\\noalign{\smallskip}
      10 &          605 &  20  & 0.0977 $\pm$ 0.0019 & 0.0971 $\pm$ 0.0019 \\\noalign{\smallskip}
      11 &          625 &  20  & 0.0959 $\pm$ 0.0025 & 0.0955 $\pm$ 0.0027 \\\noalign{\smallskip}
      12 &          645 &  20  & 0.0975 $\pm$ 0.0026 & 0.0970 $\pm$ 0.0028 \\\noalign{\smallskip}
      13 &          665 &  20  & 0.0994 $\pm$ 0.0019 & 0.0991 $\pm$ 0.0020 \\\noalign{\smallskip}
      14 &          685 &  20  & 0.1009 $\pm$ 0.0025 & 0.1006 $\pm$ 0.0026 \\\noalign{\smallskip}
      15 &          705 &  20  & 0.1009 $\pm$ 0.0024 & 0.1002 $\pm$ 0.0026 \\\noalign{\smallskip}
      16 &          725 &  20  & 0.1033 $\pm$ 0.0023 & 0.1018 $\pm$ 0.0023 \\\noalign{\smallskip}
      17 &          745 &  20  & 0.1019 $\pm$ 0.0029 & 0.1006 $\pm$ 0.0030 \\\noalign{\smallskip}
      18 &          775 &  20  & 0.0988 $\pm$ 0.0037 & 0.0973 $\pm$ 0.0043 \\\noalign{\smallskip}
      19 &          795 &  20  & 0.1019 $\pm$ 0.0020 & 0.1000 $\pm$ 0.0020 \\\noalign{\smallskip}
      20 &          815 &  20  & 0.0962 $\pm$ 0.0029 & 0.0940 $\pm$ 0.0029 \\\noalign{\smallskip}
      21 &          835 &  20  & 0.0965 $\pm$ 0.0025 & 0.0946 $\pm$ 0.0027 \\\noalign{\smallskip}
      22 &          855 &  20  & 0.0966 $\pm$ 0.0026 & 0.0948 $\pm$ 0.0030 \\\noalign{\smallskip}
      23 &          875 &  20  & 0.0981 $\pm$ 0.0040 & 0.0971 $\pm$ 0.0043 \\\noalign{\smallskip}
    \hline\noalign{\smallskip}
    \end{tabular}
\end{table}

\end{appendix}

\end{document}